\documentclass[conference,a4paper]{IEEEtran}
\usepackage[left=1.43cm,right=1.43cm,top=1.8cm,bottom=4.21cm]{geometry}
\IEEEoverridecommandlockouts
\usepackage{cite}
\usepackage{amsmath,amssymb,amsfonts,bbm}
\usepackage{algorithmic}
\usepackage{graphicx}
\usepackage{textcomp}
\usepackage[dvipsnames]{xcolor}
\usepackage{multirow}

\usepackage{subcaption}
\usepackage{caption}\usepackage{tikz}
\usepackage{tkz-tab}
\usetikzlibrary{automata,arrows,positioning,calc}
\usetikzlibrary{shapes,snakes}
\usetikzlibrary{arrows}
\usepackage{dsfont}
\usepackage{soul}
\usepackage{subfiles}
\usepackage{comment}

\def\BibTeX{{\rm B\kern-.05em{\sc i\kern-.025em b}\kern-.08em
    T\kern-.1667em\lower.7ex\hbox{E}\kern-.125emX}}

\newcommand{\blue}[1]{\textcolor{blue}{#1}}
\newcommand{\red}[1]{{\textcolor[rgb]{1,0,0}{#1}}}

\newcommand{\francesc}[1]{\noindent \blue{ {{$\blacktriangleright$ 
   {\textsf{[Francesc]: #1}} $\blacktriangleleft$}}}}

\begin{document}

\bstctlcite{IEEEexample:BSTcontrol}

\title{Throughput Analysis of IEEE 802.11bn\\Coordinated Spatial Reuse}

\author{
\IEEEauthorblockN{Francesc Wilhelmi$^{\star}$, Lorenzo Galati-Giordano$^{\star}$, Giovanni Geraci$^{\sharp}$$^{\,\flat}$,\\Boris Bellalta$^{\flat}$, Gianluca Fontanesi$^{\star}$, and David Nu\~nez$^{\flat}$ \vspace{0.1cm}
}
\IEEEauthorblockA{$^{\star}$\emph{Radio Systems Research, Nokia Bell Labs, Stuttgart, Germany}}
\IEEEauthorblockA{$^{\sharp}$\emph{Telef\'{o}nica Research, Barcelona, Spain}}
\IEEEauthorblockA{$^{\flat}$\emph{Department of Information and Communications Technologies, Universitat Pompeu Fabra, Barcelona, Spain}}
\IEEEauthorblockN{\thanks{Corresponding author: \emph{francisco.wilhelmi@nokia.com}.}
\thanks{This work was in part supported by the Spanish Research Agency through grants PID2021-123995NB-I00, PRE2019-088690, PID2021-123999OB-I00, and CEX2021-001195-M, by the UPF-Fractus Chair, and by the Spanish Ministry of Economic Affairs and Digital Transformation and the European Union through grants TSI-063000-2021-59 (RISC-6G), TSI-0630002021-138 (6G-SORUS), and TSI-063000-2021-52 (AEON-ZERO).}}
}

\maketitle

\begin{abstract}
Multi-Access Point Coordination (MAPC) is becoming the cornerstone of the IEEE 802.11bn amendment, alias Wi-Fi 8. Among the MAPC features, Coordinated Spatial Reuse (C-SR) stands as one of the most appealing due to its capability to orchestrate simultaneous access point transmissions at a low implementation complexity. In this paper, we contribute to the understanding of C-SR by introducing an analytical model based on Continuous Time Markov Chains (CTMCs) to characterize its throughput and spatial efficiency. Applying the proposed model to several network topologies, we show that C-SR opportunistically enables parallel high-quality transmissions and yields an average throughput gain of up to 59\% in comparison to the legacy 802.11 Distributed Coordination Function (DCF) and up to 42\% when compared to the 802.11ax Overlapping Basic Service Set Packet Detect (OBSS/PD) mechanism. 
\end{abstract}


\section{Introduction}
\label{sec:introduction}

As the IEEE 802.11be amendment reaches its final stages and the advent of commercial Wi-Fi 7 certified products in early 2024 approaches, the groundwork is being laid for the next phase of Wi-Fi development, IEEE 802.11bn~\cite{garcia2021ieee,khorov2020current,CheCheDas22,GalGerCar2023}. 
This new standard will pave the way for Wi-Fi~8 devices and marks a significant milestone in the evolution of Wi-Fi by targeting ultra-high reliability and opening doors to a range of challenging and emerging use cases \cite{ResCor22,OugGerPol2023}.

Before the emergence of 802.11bn, Wi-Fi focused on enhancing throughput, spectral efficiency, and reducing latency~\cite{bellalta2016ieee, lopez2019ieee, yang2020survey, deng2020ieee,CarGerKni2023}. 
However, these improvements lacked advanced coordination mechanisms among Access Points (APs). To address the issues arising from this coordination gap and enhance reliability, Wi-Fi 8 will pioneer Multi-Access Point Coordination (MAPC)~\cite{mapc}, enabling APs across different Basic Service Sets (BSSs) to explicitly collaborate and optimize spectrum resource utilization.

One notable feature within the MAPC framework is Coordinated Spatial Reuse (C-SR), which enables concurrent operations of multiple devices from distinct BSSs through adjusted transmit power management. This feature represents an improvement over its predecessor, the 802.11ax Overlapping BSS Packet Detect (OBSS/PD) SR mechanism~\cite{wilhelmi2021spatial,wilhelmi2019performance}.\footnote{For convenience, we refer to `802.11ax OBSS/PD SR' as `802.11ax SR'.} In 802.11ax SR, devices can utilize more aggressive Clear Channel Assessment (CCA) policies, albeit at the cost of limiting their power during SR-based transmission opportunities (TXOPs). Unlike 802.11ax SR, which disregards interference at the receiver, C-SR leverages the exchanged information among APs to determine the best transmit power for both concurrent transmitters, based on the anticipated signal quality of their transmissions. Furthermore, the synchronization of simultaneous transmissions within the same TXOP grants enhanced control over power adjustment.

In this paper, we introduce a new model based on Continuous Time Markov Chains (CTMCs) to analyze the throughput and spectral efficiency of IEEE 802.11bn C-SR. We further apply our model to several two-BSS topologies and explore the potential performance enhancements that C-SR offers compared to $(i)$ the legacy 802.11 Distributed Coordination Function (DCF) operation with Carrier Sense Multiple Access with Collision Avoidance (CSMA/CA) and $(ii)$ 802.11ax SR. Our main takeaways can be summarized as follows:

\begin{itemize}
    \item In all the scenarios considered, C-SR yields a higher mean throughput when compared to legacy DCF operation (up to 59\%) and 802.11ax SR (up to 42\%).
    \item When BSSs are clearly separated (e.g., with each BSS occupying a separate cubicle), C-SR provides the highest gains in terms of throughput and spectral efficiency when transmission contenders are in close proximity. This can be attributed to C-SR's capacity to facilitate parallel transmissions while controlling interference.    
    \item In more challenging deployments, like those where transmitters and receivers are randomly positioned within the same cubicle, C-SR opportunistically leverages either alternate or parallel transmissions to achieve superior throughput than both legacy DCF and 802.11ax SR.    
\end{itemize}

\section{Related Work and Contribution}
\label{sec:related_work}

Performance studies have remained a constant throughout the evolution of Wi-Fi, evaluating potential features for IEEE 802.11 amendments both analytically and through simulations. As for C-SR, initial standard contributions have demonstrated its ability to enhance throughput compared to 802.11ax SR ~\cite{csr_recap,csr_huawei,mentorBroadcom_0855r1}. The theoretical gains of C-SR when combined with Time-Division Multiple Access (TDMA) were further explored in \cite{nunez2022txop}, revealing a throughput improvement of up to 140\% in 90\% of the randomly generated scenarios. However, despite these efforts to assess its performance, uncertainty persists as to when it really is advantageous to apply C-SR.

Analytical frameworks are invaluable tools to evaluate the strengths and weaknesses of novel Wi-Fi features \cite{nguyen2007stochastic,BelCarGal2023analysis}. While Bianchi's model \cite{bianchi2000performance} stands out as one of the most widely adopted methods for characterizing Wi-Fi throughput under traffic saturation, another suitable framework are CTMCs \cite{boorstyn1987throughput}, successfully applied in prior Wi-Fi feature analysis \cite{ghaboosi2008modeling,nardelli2012closed,bellalta2016throughput,barrachina2019dynamic, barrachina2019overlap} and capable of modeling partially overlapping scenarios. This becomes particularly pertinent when studying C-SR, wherein tunable transmit power and sensitivity thresholds introduce dynamic device interactions, making it crucial to capture the variability of BSS overlap. 

In this paper, we leverage the power of CTMCs to analytically characterize the throughput of 802.11bn C-SR. Specifically, we extend and generalize the model previously proposed in \cite{wilhelmi2021spatial} for the 802.11ax SR feature and validated using the open-source Komondor simulator \cite{barrachina2019komondor}. Our work entails adapting this model to encapsulate the dynamic nature of C-SR, subsequently generating and analyzing a series of numerical results to deepen our understanding.

\section{Coordinated Spatial Reuse Procedure}
\label{sec:csr}

The MAPC feature, initially discussed during the 802.11be standardization but not implemented \cite{garcia2021ieee}, takes center stage in the forthcoming 802.11bn Task Group (TGbn). Besides C-SR, the MAPC framework may include other coordination modes such as Coordinated Beamforming (C-BF) or Coordinated Orthogonal Frequency-Division Multiple Access (C-OFDMA) \cite{GalGerCar2023}. While a concrete MAPC framework is yet to be established, AP coordination appears to revolve around two distinct roles~\cite{tgbe_compendium}: $(i)$ \textit{Sharing AP}, the winner of contention, allowing other APs to transmit during its TXOP, and $(ii)$ \textit{Shared AP}, leveraging a TXOP from a Sharing AP.

Facilitating coordination in MAPC requires the standard to introduce novel control frames and procedures for Sharing APs to communicate resource availability (e.g., indicating which frequency resources can be shared and their duration) and for Shared APs to specify requirements (e.g., necessary OFDMA resource sets). As a reference in this paper for modeling the C-SR feature, we consider the following procedure:

\begin{enumerate}
    \item \emph{TXOP winning:} Devices engage in Listen-Before-Talk (LBT)-based channel access and the AP with the lowest random backoff becomes the Sharing AP.
    \item \emph{Information collection:} APs gather information to derive potential groups of simultaneous transmissions.
    \item \emph{Shared TXOP definition:} The Sharing AP determines its next transmission based on the potential set of Shared APs and their impact on the targeted Signal-to-Interference-plus-Noise Ratio (SINR). The Sharing AP may impose transmission constraints on Shared APs.
    \item \emph{Coordination trigger:} Via a trigger frame, the Sharing AP invites the Shared APs to participate in the upcoming shared TXOP. The trigger frame outlines the rules of the transmission, including the maximum power for Shared AP(s) or associated stations (STAs). The Sharing AP decides which devices benefit from the shared TXOP.
    \item \emph{Coordinated data transmission:} Transmission takes place, adhering to the power limitations imposed. Shared AP(s) determine the optimal STAs and Modulation and Coding Scheme (MCS) for transmission.
\end{enumerate}

The definition of the shared TXOP may significantly affect the efficacy of C-SR, with various possible objectives based on the use case. In this paper, we assume that the Sharing AP employs maximum transmit power, while the Shared APs limit their power so as to ensure that the SINR at the STA associated with the Sharing AP does not fall below a predefined Capture Effect (CE) threshold, and therefore, that simultaneous transmissions remain successful. Moreover, a detailed definition of the coordination mechanism, including the necessary signaling overheads, is currently under discussion and remains beyond the scope of this paper. 

\section{Throughput Analysis via CTMCs}
\label{sec:model}

A CTMC represents the behavior of a system as a stochastic process, allowing the calculation of its steady-state performance \cite{boorstyn1987throughput}. In characterizing Wi-Fi through CTMCs, we assume two conditions: $(i)$ a state $s\in \mathcal{S}$ is defined by a set $\Phi_s$ of active BSSs, and $(ii)$ stochastic processes govern channel access contention and transmission durations, both adhering to exponential distributions. For the sake of clarity, this paper considers solely downlink transmissions and BSSs comprising a single AP and STA. However, the proposed model can accommodate multiple transmitting pairs and uplink transmissions, albeit leading to an increased number of states.

Deriving the stationary distribution of a CTMC involves solving $\vec{\pi} Q = 0$, where $\vec{\pi}$ constitutes an array of probabilities associated with each potential chain state and $Q$ represents the transitions probability matrix, with $Q_{i,j}$ denoting the transition from state $i$ to $j$. The rates of forward and backward transitions are denoted as $\lambda$ and $\mu$, respectively. In the following, we outline how transition rates are computed, contingent upon the chosen channel access policies and the models' governing packet transmission durations.

\subsection{CTMC Representation}
\label{sec:ctmc_example}

\begin{figure}[t!]
    \centering    \includegraphics[width=\columnwidth]{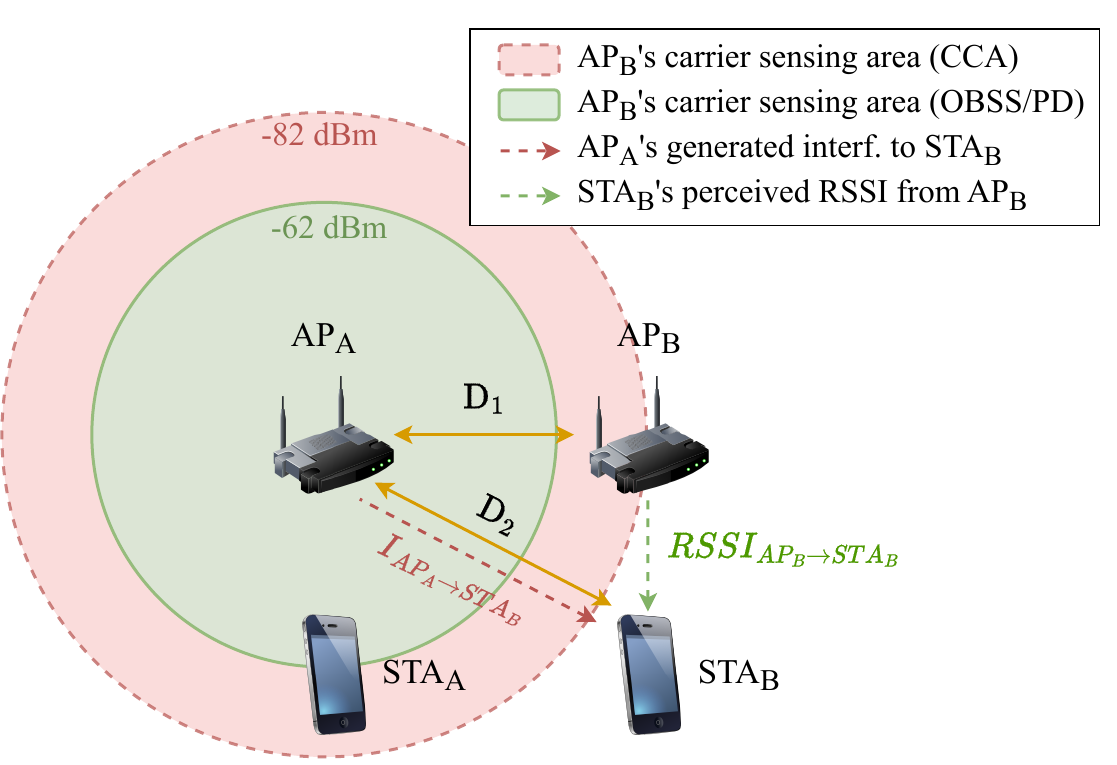}    \caption{Distribution of nodes and their interactions in the two-BSS toy scenario, used to illustrate the CTMC construction.}    \label{fig:ctmn_scenario}
\end{figure}

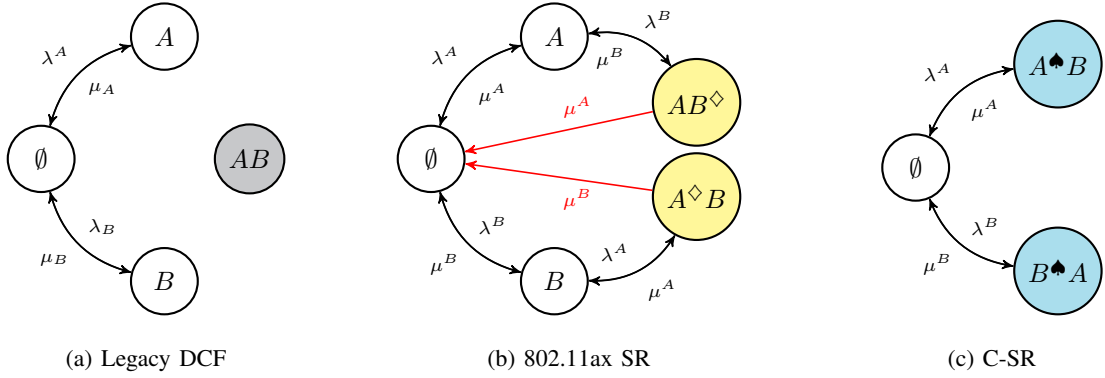
\begin{figure*}[t!]
    \centering
    \begin{subfigure}{0.3\textwidth}
        \begin{center}
        \begin{tikzpicture}[->, >=stealth', auto, semithick, node distance=3cm]
        \tikzstyle{every state}=[fill=white,draw=black,thick,text=black,scale=1]
        \node[state]    (0)                     {$\emptyset$};
        \node[state]    (A)[above right of=0,xshift=-0.5cm, yshift=-0.5cm]   {$A$};
        \node[state]    (B)[below right of=0, xshift=-0.5cm, yshift=0.5cm]   {$B$};
        \node[state]    (AB)[below right of=A, xshift=-1cm, yshift=0.5cm, fill=Gray!50]   {$AB$};
        \path
        (0) edge[bend left] node{\scriptsize$\lambda^A$} (A)
            edge[bend right] node{\scriptsize$\lambda_B$}  (B)
        (A) edge[bend right] node{\scriptsize$\mu_A$} (0)
        (B) edge[bend left] node{\scriptsize$\mu_B$} (0);
        \end{tikzpicture}
        \end{center}   
        \caption{Legacy DCF}
        \label{fig:ctmn_csma}
    \end{subfigure}
    \begin{subfigure}{0.3\textwidth}        
        \begin{center}
        \begin{tikzpicture}[->, >=stealth', auto, semithick, node distance=3cm]
        \tikzstyle{every state}=[fill=white,draw=black,thick,text=black,scale=1]
        \node[state]    (0)    {$\emptyset$};
        \node[state]    (A)[above right of=0, xshift=-0.5cm, yshift=-0.5cm]   {$A$};
        \node[state]    (B)[below right of=0,  xshift=-0.5cm, yshift=0.5cm]   {$B$};
        \node[state] (AB)[below right of=A, xshift=-0.25cm, yshift=1.25cm, fill=yellow!50] {$AB^\diamondsuit$};
        \node[state] (BA)[below of=AB, yshift=1.75cm, fill=yellow!50] {$A^\diamondsuit B$};
        \path
        (0) edge[bend left] node{\scriptsize$\lambda^A$} (A)
            edge[bend right] node{\scriptsize$\lambda^B$}  (B)
        (A) edge[bend right] node{\scriptsize$\mu^A$} (0)
        (B) edge[bend left] node{\scriptsize$\mu^B$} (0)
        (A) edge[bend left] node{\scriptsize$\lambda^B$} (AB)
        (B) edge[bend right] node{\scriptsize$\lambda^A$} (BA)   
        (AB) edge[bend right] node{\scriptsize$\mu^B$} (A)
        (BA) edge[bend left] node{\scriptsize$\mu^A$} (B)
        (AB) edge[right, color=red, pos=0.4, above] node{\scriptsize$\mu^A$} (0)
        (BA) edge[left, color=red, pos=0.4, below] node{\scriptsize$\mu^B$} (0)        
        ;
        \end{tikzpicture}
        \end{center}        
        \caption{802.11ax SR}
        \label{fig:ctmn_11ax_sr}
    \end{subfigure}
    \begin{subfigure}{0.3\textwidth}
        \begin{center}
        \begin{tikzpicture}[->, >=stealth', auto, semithick, node distance=3cm]
        \tikzstyle{every state}=[fill=white,draw=black,thick,text=black,scale=1]
        \node[state] (0)              {$\emptyset$};
        \node[state] (A)[above right of=0, xshift=-0.25cm, yshift=-0.75cm,
        fill=SkyBlue!50]   {$A^\spadesuit B$};
        \node[state] (B)[below right of=0, xshift=-0.25cm, yshift=0.75cm,
        fill=SkyBlue!50]   {$B^\spadesuit A$};
        \path
        (0) edge[bend left] node{\scriptsize$\lambda^A$} (A)
            edge[bend right] node{\scriptsize$\lambda^B$}  (B)
        (A) edge[bend right] node{\scriptsize$\mu^A$} (0)
        (B) edge[bend left] node{\scriptsize$\mu^B$} (0)
        ;
        \end{tikzpicture}
        \end{center}        
        \caption{C-SR}
        \label{fig:ctmn_csr}
    \end{subfigure}
    \caption{CTMC representation of the two-BSS toy scenario: (a) Legacy DCF, (b) 802.11ax SR, and (c) C-SR. Gray coloring designates infeasible states, while yellow and blue signify states in which two BSSs access the channel through 802.11ax SR and C-SR, respectively. Diamond superscripts denote the use of 802.11ax SR TXOP for transmission, while spades indicate the Sharing AP in a C-SR TXOP.}
    \label{fig:ctmn}
\end{figure*}

Fig.~\ref{fig:ctmn_scenario} shows a deployment referred to as the \emph{two-BSS toy scenario}. This deployment serves as an illustrative tool to characterize the different channel access modes---legacy DCF, 802.11ax SR, and C-SR---using CTMC. The scenario involves two BSSs, named as BSS $A$ and BSS $B$, each comprising an AP and a STA. Considering downlink transmissions and the PHY layer abstraction detailed in the sequel as per \eqref{eq:pathloss}, the separation between APs, $D_1$, governs channel access for legacy DCF and 802.11ax SR. Similarly, the distance between a given AP and the STA of the opposing BSS, $D_2$, determines the mutual interference when both APs transmit simultaneously.

To illustrate the resulting CTMC of the configuration in Fig.~\ref{fig:ctmn_scenario} when applying each channel access mode, we account for the following scenario-dependent conditions: $(i)$ the power sensed by transmitters in either BSS $A$ or $B$ surpasses the $-82$\,dBm CCA threshold when the other accesses the channel, $(ii)$ the power sensed by transmitters in either BSS $A$ or $B$ falls below the $-62$\,dBm 802.11ax OBSS/PD SR threshold when the other accesses the channel, and $(iii)$ coordinated transmissions become feasible since the SINR at both BSS $A$ and $B$ receivers exceeds the CE threshold for the agreed configuration. Based on these assumptions, Fig.~\ref{fig:ctmn} illustrates the resulting CMTC for each approach. Note that, while the node locations and propagation effects remain consistent across all CTMC, the set of possible states varies based on the chosen channel access mechanism, as detailed next.

\subsubsection*{Legacy DCF operation} 
As depicted in Fig.~\ref{fig:ctmn_csma}, only transitions to isolated states are viable, i.e., either BSS $A$ or $B$ transmitting independently. In other words, BSS $A$ or $B$ can only transmit individually if the interference they introduce to the other surpasses the CCA threshold. To point this out, we have included the unfeasible state $AB$ within the CTMC (highlighted in gray) in Fig.~\ref{fig:ctmn_csma}. In this state, both BSSs $A$ and $B$ would transmit simultaneously, an occurrence that contradicts the established channel access policy. Hence, such state cannot be transitioned to from any other state, and its steady-state probability is zero.

\subsubsection*{802.11ax SR} 
As shown in Fig.~\ref{fig:ctmn_11ax_sr}, each BSS maintains the ability to access the channel akin to the legacy mode. However, when one BSS seizes the channel, the other can also transmit with constrained power. Within Fig.~\ref{fig:ctmn_11ax_sr}, states involving the utilization of an SR TXOP are shown in yellow and denoted as $AB^\diamondsuit$ and $A^\diamondsuit B$. The device engaging in an SR TXOP, which enforces a transmit power limitation as a linear function of the selected 802.11ax OBSS/PD SR threshold~\cite{wilhelmi2021spatial}, is indicated by a superscript diamond. 
Note that forward transitions into SR TXOP states ($A^\diamondsuit B$ and $AB^\diamondsuit$) are not possible from state~$\emptyset$, since SR TXOPs are detected once the contention winner initiates transmission. Backward transitions from SR TXOP states are indicated in red within the CTMC.

\subsubsection*{Coordinated spatial reuse}  
C-SR empowers both BSSs to transmit concurrently, as exemplified in Fig.~\ref{fig:ctmn_csr}. The contention winner determines the maximum allowed transmit power for the other BSS, ensuring that newly generated interference does not excessively degrade the SINR at the STA of the sharing BSS. In Fig.~\ref{fig:ctmn_csr}, states involving C-SR TXOPs are indicated in blue, with a spade denoting the sharing AP (e.g., $A^\spadesuit$). Importantly, the finalization of a C-SR state hinges on the sharing AP, causing the exclusive transition to the empty state, $\emptyset$. This transition aligns with the Sharing AP's TXOP duration, ruled by $\mu$.

\subsection{Channel Access Model, $\lambda$}
\label{sec:lambda}

The initiation of transmissions depends on the adopted channel access mechanism and on the impact of neighboring devices in terms of induced interference. In all scenarios, a contention phase is employed to determine the winner of the upcoming TXOP. Each device independently decrements a random backoff counter as long as the channel is detected as idle. A device deems the channel as idle if the sensed power remains below a defined channel-clear threshold (CCA or 802.11ax SR threshold, depending on the case). 

The sensed power at device $n$, $P^\text{RX}_n$ (dBm), is influenced by ongoing transmissions from a set of interfering devices $\mathcal{I}$ and is expressed as:
\begin{equation}
    P^\text{RX}_n = P^\text{Noise} + \sum_{m \in \mathcal{I}} P_m^\text{TX} + G^\text{TX} + G^\text{RX} - \text{PL}(d_{n,m}),
    \label{eq:pathloss}
\end{equation}
where $P^\text{Noise}$ represents the noise power, $P_m^\text{TX}$ is the transmit power employed by interfering device $m$, and $G^\text{TX/RX}$ denotes the antenna gain at the transmitter/receiver. The path loss $\text{PL}(d_{n,m})$ (dB) between devices $n$ and $m$ follows a standard log-distance model incorporating shadowing effects (see Table~\ref{tbl:simulation_parameters} for additional details):
\begin{equation}
    \text{PL}(d_{n,m}) = \text{PL}_{0} + 10\alpha \log_{10}(d_{n,m}) + \frac{\sigma}{2} + \frac{\omega}{2} \frac{d_{n,m}}{10}.
    \label{eq:pathloss2}
\end{equation}

Irrespective of the channel access mode (that determines the feasibility of state transitions), the attempt rate is defined as the reciprocal of the expected backoff time, given by
\begin{equation}
    \lambda = \frac{1}{E[\text{backoff}]} = \frac{2}{\text{CW}-1} \, T_{\textrm{e}},
    \label{eq:lambda}
\end{equation}
where $T_{\textrm{e}}$ is the duration of an empty slot (9 $\mu$s). In the context of C-SR, $\lambda$ governs the channel access rate of the TXOP winner. However, even if a Shared AP is not the TXOP winner, it can still access the medium when permitted by the Sharing AP through explicit TXOP sharing. In the CTMC model, whenever a coordinated transmission is enabled by C-SR, both Sharing and Shared APs access the channel simultaneously, based on the Sharing AP's $\lambda$. For instance, as illustrated in Fig.~\ref{fig:ctmn_csr}, $\lambda^A$ is the transition probability to reach state $A^\spadesuit B$.

\subsection{Transmission Duration Model, $\mu$}
\label{sec:mu}

The transmission duration $1/\mu$ indicates for how long a device occupies the channel after gaining access. We differentiate between two durations, depending upon the success or failure of the data transmissions in a given state. We employ the frames and intervals specified by the IEEE 802.11 protocol~\cite[Table~B.6]{wilhelmi2021spatial} to quantify both successful ($T_{\mathrm{s}}$) and unsuccessful ($T_{\mathrm{u}}$) channel access durations. In particular,
\begin{equation}
   \begin{cases}
       T_{\mathrm{s}} = & T_\text{RTS} + 3\cdot T_\text{SIFS} + T_\text{CTS} + T_\text{DATA} + \\&T_\text{ACK} + T_\text{DIFS} + T_{\mathrm{e}} \\
        T_{\mathrm{u}} = & T_\text{RTS} + T_\text{SIFS} + T_\text{CTS} + T_\text{DIFS} + T_{\mathrm{e}}.
   \end{cases}
   \label{eq:successful_unsuccessful_slots}
\end{equation}

While the duration of control frames (e.g., acknowledgements) remains constant, the length of a data transmission depends on the number of aggregated MAC Service Data Unit (MSDU) frames and the MCS used. 
Based on the SINR $\gamma_i^{(n)}$ at BSS $n$ during its transmission in state $i$, the departure rate from state $i$ to $j$, $\mu_{i,j}$ is given by
\begin{equation}
    \mu_{i,j} = \begin{cases}
        1/T_{\mathrm{s}}, & \text{if } n\in \Phi_i  \land \Phi_i\setminus n = j \land \gamma^{(n)}_{i} \geq \gamma_{\textrm{CE}} \\
        1/T_{\mathrm{u}}, & \text{if } n\in \Phi_i  \land \Phi_i\setminus n = j \land \gamma^{(n)}_{i} < \gamma_{\textrm{CE}}\\
        0, & \text{otherwise }.
    \end{cases}
    \label{eq:mu}
\end{equation}

For a transmission by BSS $n$ to be successful from a given state $i$, it is required that $\gamma_i^{(n)}$ surpasses a predefined CE threshold, $\gamma_{\textrm{CE}}$. Notice as well that this condition strongly depends on the set of concurrent transmissions within the same state, $\Phi_i$, a set determined by the channel access policy employed, as described in earlier subsections. 

\subsection{Key Performance Indicators}
\label{sec:throughput_reliability}

The construction of a CTMC allows for computing the steady-state performance for the characterized OBSS. In this paper, we consider three performance metrics described next.

\subsubsection*{Throughput} 
The throughput $\Gamma^{(n)}$ (bps) of BSS $n$ is defined as
\begin{equation}
    \Gamma^{(n)} = \sum_{s \in \mathcal{S}} \pi_s \mathbbm{1}_{\{\gamma^{(n)}_s \geq \gamma_{\textrm{CE}}\}}  \big( N^{(n)}_{a,s} \, L_D \, R_s^{(n)} \big),
    \label{eq:throughput}
\end{equation}
where $\mathbbm{1}_{\{\cdot\}}$ denotes the indicator function and $N^{(n)}_{a,s}$ is the number of aggregated MSDUs transmitted by BSS $n$ in state $s$ (capped by the maximum TXOP duration). $L_D$ represents the length of an individual data payload frame and $R_s^{(n)}$ denotes the rate employed by BSS $n$ in state $s$. The rate depends on the AP's transmit power and separation from its STA.

\subsubsection*{Airtime} 
The airtime of a BSS is calculated as the sum of steady-state probabilities linked to states in which the BSS remains active. Unlike legacy DCF, both 802.11ax SR and C-SR strive to augment airtime by identifying extra TXOPs through concurrent transmissions. The airtime $\rho^{(n)}\in[0,100]$ of BSS $n$ is expressed as
\begin{equation}
    \rho^{(n)} = 100 \cdot \sum_{s \in \mathcal{S}} \pi_s \mathbbm{1}_{\{n\in \Phi_s\}}.
    \label{eq:airtime}
\end{equation}

\subsubsection*{Spatial efficiency}
For a more profound comprehension of the impact of implementing C-SR, a spatial efficiency metric $\zeta^{(n)}\in[0,1]$ is introduced as
\begin{equation}
    \zeta^{(n)} = \sum_{s \in \mathcal{S}} \pi_s \tau_s^{(n)},
    \label{eq:reliability}
\end{equation}
where $\tau_s^{(n)}$ is the ratio between the achieved throughput and the data rate employed by BSS $n$ in state $s$. Specifically, $\tau_s^{(n)} = \Gamma_s^{(n)}/R_s^{(n)}$. Intuitively, spatial efficiency reflects how effectively the spectrum is used by combining aspects like contention, transmission failures, and the quality of transmissions. Maximum spatial efficiency $\zeta^{(n)}=1$ is achieved when the devices in BSS $n$ can access the medium when needed and perform successful transmissions.

\captionsetup[subfigure]{labelformat=empty}
\begin{figure*}[!t]
\centering
\subfloat[Settings 1a and 1b]{\includegraphics
[height=50mm]
{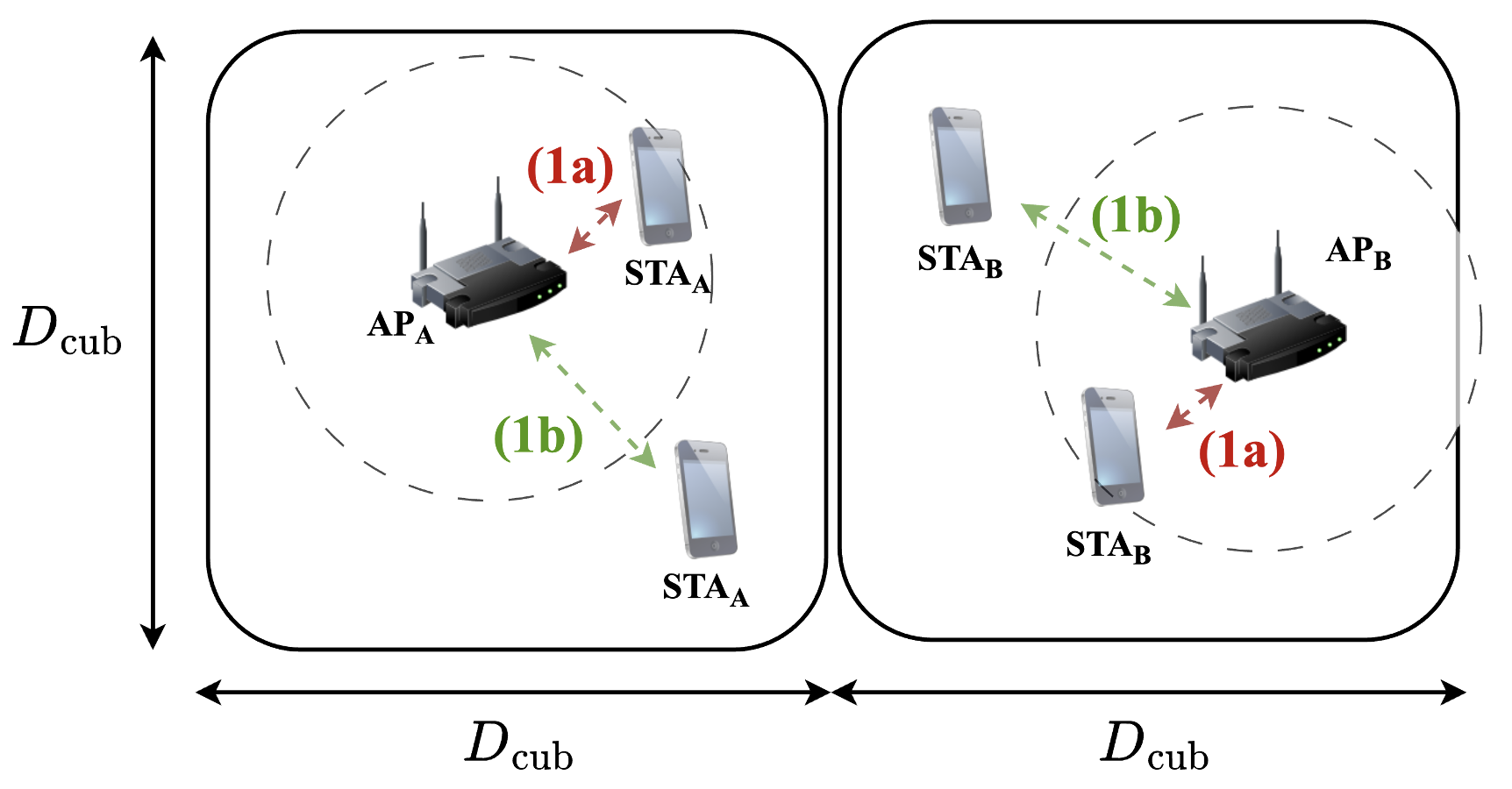}
\label{fig:setting1}}
\hspace*{18.0mm}
\subfloat[Settings 2a and 2b]{\includegraphics
[height=50mm]
{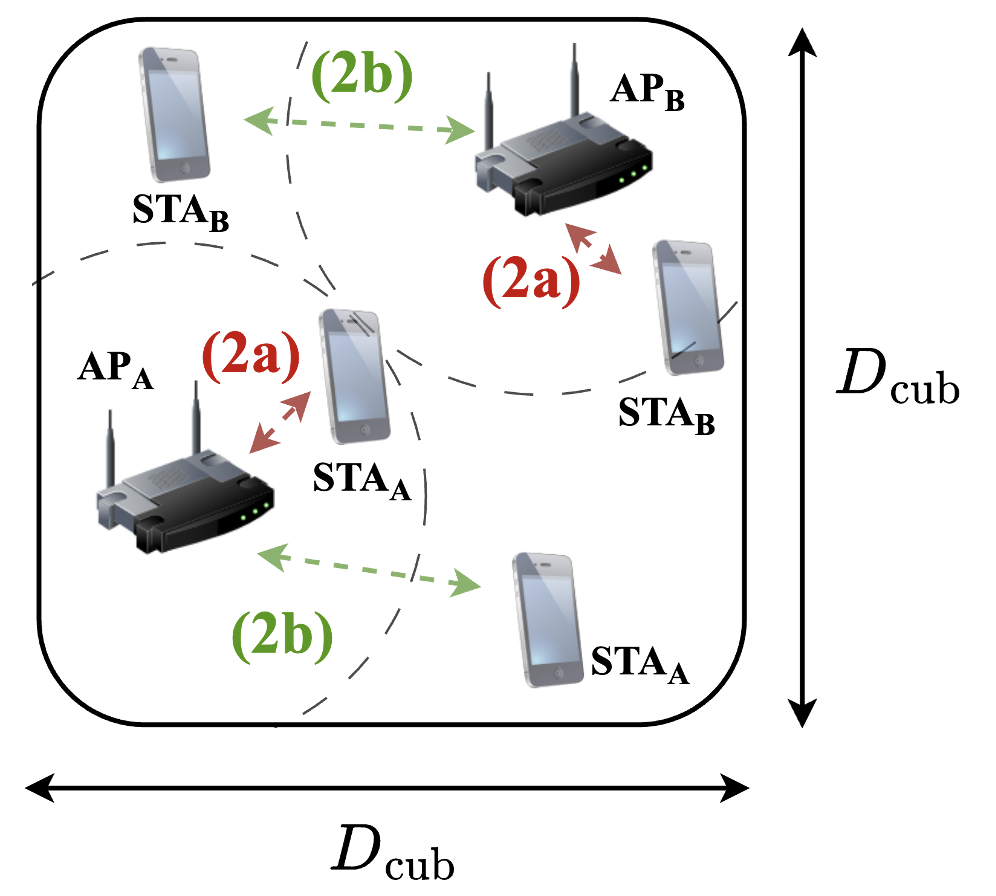}
\label{fig:setting2}}
\captionsetup{justification=justified}
\caption{Illustration of the four settings considered. Left: different cubicles with either nearby STAs [Setting 1a] or unconstrained STAs [Setting 1b]. Right: same cubicle with either nearby STAs [Setting 2a] or unconstrained STAs [Setting 2b].}
\label{fig:settings}
\end{figure*}

\section{Numerical Results and Discussion}
\label{sec:performance_evaluation}

To evaluate the performance gains achieved by C-SR when compared to both legacy DCF and 802.11ax SR, we investigate the scenarios illustrated in Fig.~\ref{fig:settings}. Specifically, we explore four distinct settings:
\begin{itemize}
   \item{\emph{[Setting 1a] Different cubicles, nearby STAs:}} 
APs are distributed randomly across separate and contiguous square-shaped cubicles, each of size $D_{\textrm{cub}}$. The maximum separation between an AP and its associated STA, located within the same cubicle as the AP, is fixed to $D_{\textrm{AP-STA}}^{\max} = 2$\,m.
    \item{\emph{[Setting 1b] Different cubicles, unconstrained STAs:}}
APs are still situated in separate cubicles, but STAs are now positioned randomly anywhere within the same cubicle as their respective AP.
\item{\emph{[Setting 2a] Same cubicle, nearby STAs:}} 
Here, both APs and STAs are now randomly placed within a single cubicle of size $D_{\textrm{cub}}$. STAs are restricted to a maximum distance of $D_{\textrm{AP-STA}}^{\max} = 2$\,m from their associated AP.
\item{\emph{[Setting 2b] Same cubicle, unconstrained STAs:}} 
For this setting, both APs and STAs are arbitrarily distributed within a single shared cubicle, devoid of any constraints regarding minimum distances between APs or between the APs and their respective STAs.
\end{itemize}

\begin{figure*}[t!]
    \centering
    \includegraphics[width=\textwidth]{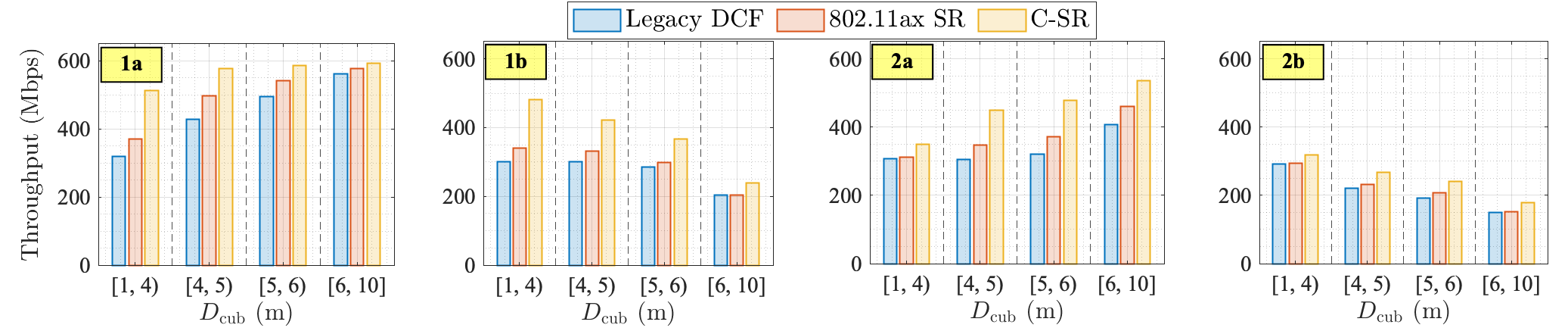}
    \caption{Mean throughput experienced by the BSSs vs. cubicle size $D_{\textrm{cub}}$ for each approach across the four settings analyzed. Mean is computed for realizations arranged in different $D_{\textrm{cub}}$ value intervals: $[1, 4)$\,m, $[4, 5)$\,m, $[5, 6)$\,m, and $[6, 10]$\,m.}
    \label{fig:results_random}
\end{figure*}

\begin{figure*}[t!]
    \centering
    \includegraphics[width=\textwidth]{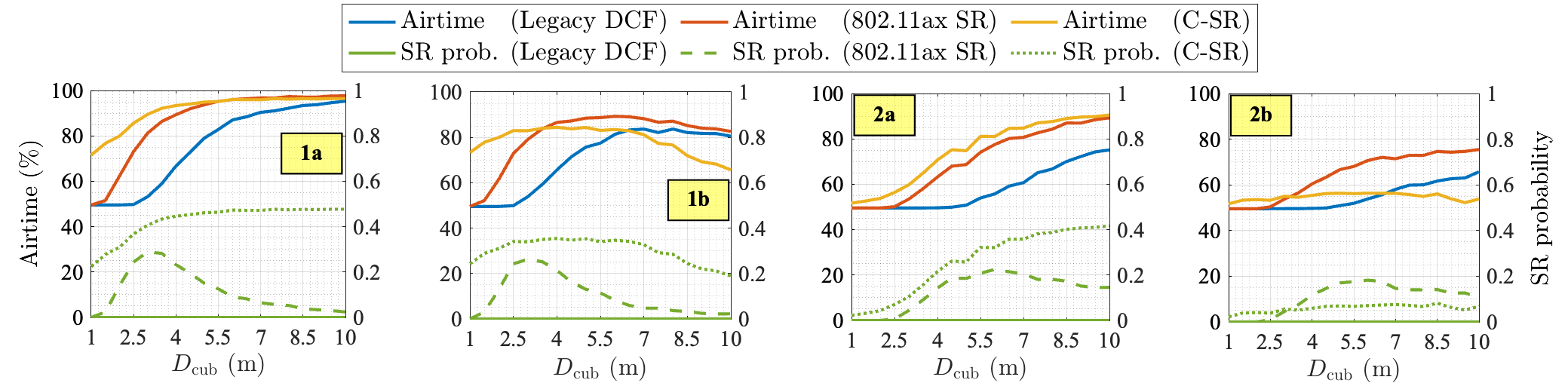}
    \caption{Mean airtime experienced by the BSSs vs. cubicle size $D_{\textrm{cub}}$ for each approach across the four settings analyzed (left y-axis). The probability for a BSS to employ a SR transmission opportunity is also shown in green (right y-axis). 
    }
    \label{fig:txop_results}
\end{figure*}

All values of $D_{\textrm{cub}} \in [1, 10]$\,m are tested with a 0.1\,m step. For each value, the four settings are generated randomly and evaluated through the proposed CTMC framework 1000 times.
Table~\ref{tbl:simulation_parameters} details the system parameters employed.\footnote{As we aim to determine the upper bound gains offered by C-SR, we neglect the communication overheads required for enabling the coordination between APs, which are envisioned to be minimal with respect to other MAPC features \cite{mentorLG_0854r0}. 
However, these overheads can be accounted for in the CTMC model by adjusting the transmission duration model detailed in Section~\ref{sec:mu}.}
For each setting and channel access mechanism, the results presented in the remainder of this section include:
\begin{itemize}
    \item \emph{Mean throughput:} Computed from \eqref{eq:throughput}, shown in Fig.~\ref{fig:results_random}.
    \item \emph{Airtime:} Obtained as per \eqref{eq:airtime} and shown in Fig.~\ref{fig:txop_results} along with the probability for a BSS to employ an SR TXOP.
    %
    \item \emph{Spatial efficiency:} 
    In line with \eqref{eq:reliability} and plotted in Fig.~\ref{fig:allsettings_sr_gain}.
\end{itemize}

\begin{table}[t!]
\centering
\caption{System parameters employed.}
\label{tbl:simulation_parameters}
\resizebox{\columnwidth}{!}{%
\begin{tabular}{|c|l|c|}
\hline
 \multicolumn{1}{|c|}{\textbf{Parameter}} & \multicolumn{1}{c|}{\textbf{Description}} & \multicolumn{1}{c|}{\textbf{Value}}\\ \hline
$B$ & Transmission bandwidth & 80 MHz \\ \cline{1-3} 
$F_c$ & Carrier frequency & 6 GHz\\ \cline{1-3}
$SUSS$ & Single-user spatial streams & 2 \\ \cline{1-3} 
$P^\text{TX}$ & Transmit power & 20 dBm \\ \cline{1-3} 
$P^\text{Noise}$ & Noise power & -95 dBm\\ \cline{1-3} 
$G^\text{TX/RX}$ & Transmitter/receiver antenna gain & 0/0 dB \\ \cline{1-3} 
 $PL_0$ & Loss at the reference dist. & 5 dB \\ \cline{1-3} 
$\alpha$ & Path-loss exponent & 4.4 \\ \cline{1-3} 
$\sigma$ & Shadowing factor & 9.5 \\ \cline{1-3} 
$\omega$ & Obstacles factor & 30 \\ \cline{1-3} 
$\gamma_{\textrm{CE}}$ & Capture effect threshold & 10 dB\\ \cline{1-3} 
$\text{CW}_{\min/\max}$ & Min/max contention window & 32 \\ \cline{1-3} 
$L_{D}$ & Length of data packets & 12000 bits \\
\hline
\end{tabular}%
}
\end{table}



\subsection{Throughput and Airtime}

\subsubsection*{[Setting 1a] Different cubicles, nearby STAs} 
Beginning with the left-most plot in Fig.~\ref{fig:results_random}, this scenario proves to be where C-SR exhibits its most prominent gains as the conditions---characterized by nearby STAs and appropriately spaced APs---are well suited for coordination. C-SR identifies and utilizes a greater number of parallel transmissions compared to other approaches, outperforming legacy DCF and 802.11ax SR by 59\% and 38\%, respectively, for $D_{\textrm{cub}} \in [1, 4)$\,m. 
This is corroborated by the airtime and the probability of SR opportunities (left-most plot in Fig.~\ref{fig:txop_results}) achieved by C-SR, especially in situations of high contention, such as when $D_{\textrm{cub}}$ is low. As $D_{\textrm{cub}}$ increases, contention effects become less pronounced, reducing the necessity for coordinated transmissions. For $D_{\textrm{cub}} \in [6, 10]$\,m, C-SR's throughput gains over legacy DCF and 802.11ax SR diminish to 6\% and 3\%, respectively.

\subsubsection*{[Setting 1b] Different cubicles, unconstrained STAs} 
The second plots in Fig.~\ref{fig:results_random} and Fig.~\ref{fig:txop_results} refer to an unconstrained positioning of STAs, which at greater distances (as $D_{\text{cub}}$ increases) results in SINR degradation. For $D_{\textrm{cub}} \in [1, 4)$\,m, C-SR allows an increasingly higher number of parallel transmissions and throughput, the latter with improvements of 59\% and 42\% over legacy DCF and 802.11ax SR, respectively. Nevertheless, beyond $D_{\textrm{cub}}=5$\,m, the airtime achieved by C-SR diminishes progressively with increasing distance since the probability of successful parallel transmissions decreases as $D_{\textrm{cub}}$ rises. In such cases, although the substantial separation between transmitters favors coordinated transmissions, the SINR at STAs remains low due to their distance from their respective APs. C-SR effectively adapts to this trend and sometimes prioritizes alternating transmissions over parallelization to attain improved SINR.

\subsubsection*{[Setting 2a] Same cubicle, nearby STAs}
Much like in Setting~1a, C-SR significantly outperforms other methods in the third plot of Fig.~\ref{fig:results_random}. However, Setting~2a presents a more challenging scenario due to the closer proximity of APs. This proximity makes C-SR more cautious regarding parallel transmissions, as can be observed from the achieved airtime and probability of SR opportunity (third plot in Fig.~\ref{fig:txop_results}). Despite 802.11ax SR and C-SR achieving similar airtime, C-SR's configurations are less conservative (it detects more SR opportunities) than those of 802.11ax SR and result in higher throughput. Unlike Setting~1a, where the greatest gains were observed at shorter distances, C-SR yields superior throughput gains in Setting~2a at mid-range distances compared to legacy DCF and 802.11ax SR (about 50\% and 30\%, respectively).

\subsubsection*{[Setting 2b] Same cubicle, unconstrained STAs} 
This scenario, characterized by random placement of both APs and STAs across the cubicle, poses the most demanding conditions in terms of contention and interference. Consequently, the advantages of C-SR are reduced in most instances (fourth plot in Fig.~\ref{fig:results_random}). Here, both contention and inter-AP interference remain high for low $D_{\textrm{cub}}$ values, preventing all mechanisms from simultaneously transmitting. Moreover, as $D_{\textrm{cub}}$ increases, contention eases, but at the cost of reduced SINR, undermining the quality of potential parallel transmissions. In this context, C-SR adopts a more cautious approach to channel access, as evident from the resulting airtime (fourth plot in Fig.~\ref{fig:txop_results}). For instance, for $D_{\textrm{cub}} > 6$\,m even legacy DCF achieves greater airtime. This is attributed to C-SR's avoidance of parallel transmissions with poor SINR, even when they are feasible according to CCA conditions. This behavior, reaffirmed by the SR opportunities detected by C-SR, leads to throughput improvements of up to 19\% and 17\% compared to legacy DCF and 802.11ax SR, respectively.

\subsection{Spatial Efficiency}

\begin{figure}[t!]
    \centering    \includegraphics[width=.99\columnwidth]{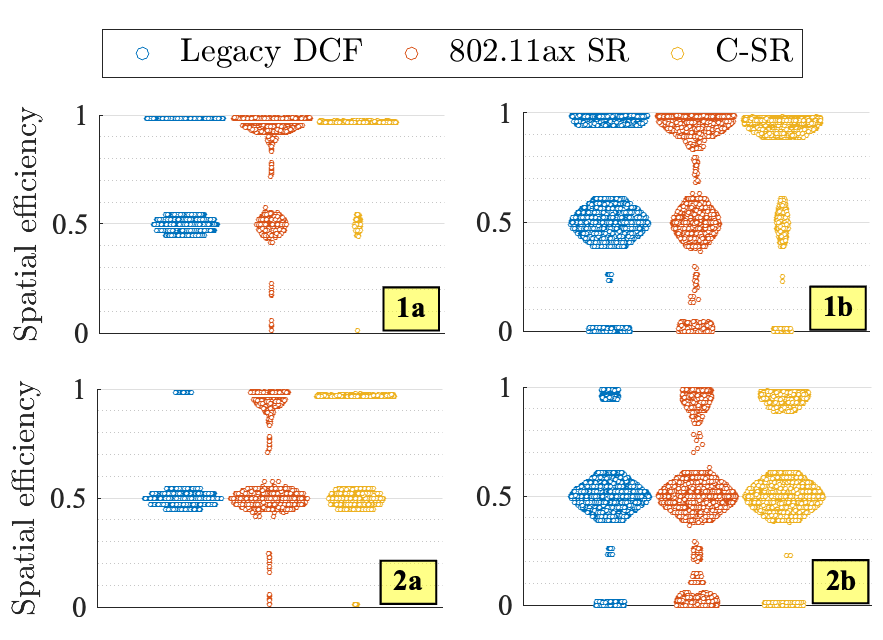}
    \caption{Distribution of spatial efficiency observations for each approach across the four settings analyzed.}
    \label{fig:allsettings_sr_gain}
\end{figure}

We now focus on the spatial efficiency achieved by each method. To evaluate this, Fig.~\ref{fig:allsettings_sr_gain} presents a swarm plot for each setting, depicting the distribution of spatial efficiency observations. Each distribution extends along the x-axis, with varying concentrations of specific values.

Across all cases, C-SR consistently outperforms the other mechanisms in terms of spatial efficiency. In particular, C-SR is shown to be very effective in Settings 1a, 1b, and 2a. In Setting 2b, the C-SR's gains in terms of spatial efficiency appear less evident. However, as demonstrated earlier, C-SR still improves over legacy DCF and 802.11ax SR by identifying situations where simultaneous transmissions are feasible with a certain minimum SINR, versus cases where it is more advantageous to alternate transmissions.

Finally, we observe that for certain deployments enabling SR (either via 802.11ax SR or C-SR) does not pay off, as the increased interference outweighs the benefits of concurrent transmissions. The balance between these two factors is heavily influenced by the channel-clear thresholds employed by each method. A comprehensive study and optimal dimensioning of such thresholds will be crucial in achieving optimal spatial efficiency.

\section{Conclusions}
\label{sec:conclusions}

As we stand at the beginning of the IEEE 802.11bn standardization (Wi-Fi 8), it becomes imperative to assess the performance of candidate features, understanding their limitations and scope for enhancements. In this paper, we introduced an analytical model based on Markov chains to characterize the performance of C-SR, a promising feature aimed at enhancing spectral efficiency through coordination.

Our findings showed that C-SR enhances the mean throughput compared to legacy DCF and 802.11ax SR, also exhibiting superior spatial efficiency. Unlike its predecessor, 802.11ax SR, C-SR identifies and capitalizes on transmission opportunities, enabling multiple devices to successfully transmit at the same time through coordinated efforts. This quality is particularly appealing, as it mitigates the risk of unsuccessful transmissions, which may result from employing more aggressive channel access policies like 802.11ax SR. Indeed, the latter relies solely on the power sensed at transmitters without accounting for the interference perceived at the receivers.

To fully comprehend the potential advantages and limitations of C-SR, future research could extend our evaluation to include larger-scale deployments, including dense residential and enterprise environments. Additionally, investigating the performance of C-SR in the presence of legacy lacking coordination capabilities would offer valuable insights for its practical implementation.
%
%

\bibliographystyle{IEEEtran}
\bibliography{bib}

\begin{thebibliography}{10}
\providecommand{\url}[1]{#1}
\csname url@samestyle\endcsname
\providecommand{\newblock}{\relax}
\providecommand{\bibinfo}[2]{#2}
\providecommand{\BIBentrySTDinterwordspacing}{\spaceskip=0pt\relax}
\providecommand{\BIBentryALTinterwordstretchfactor}{4}
\providecommand{\BIBentryALTinterwordspacing}{\spaceskip=\fontdimen2\font plus
\BIBentryALTinterwordstretchfactor\fontdimen3\font minus
  \fontdimen4\font\relax}
\providecommand{\BIBforeignlanguage}[2]{{%
\expandafter\ifx\csname l@#1\endcsname\relax
\typeout{** WARNING: IEEEtran.bst: No hyphenation pattern has been}%
\typeout{** loaded for the language `#1'. Using the pattern for}%
\typeout{** the default language instead.}%
\else
\language=\csname l@#1\endcsname
\fi
#2}}
\providecommand{\BIBdecl}{\relax}
\BIBdecl

\bibitem{garcia2021ieee}
A.~Garcia-Rodriguez, D.~L\'{o}pez-P\'{e}rez, L.~Galati-Giordano, and G.~Geraci,
  ``{IEEE 802.11be: Wi-Fi 7 strikes back},'' \emph{{IEEE Communications
  Magazine}}, vol.~59, no.~4, pp. 102--108, 2021.

\bibitem{khorov2020current}
E.~Khorov, I.~Levitsky, and I.~F. Akyildiz, ``{Current status and directions of
  IEEE 802.11be, the future Wi-Fi 7},'' \emph{{IEEE Access}}, vol.~8, pp.
  88\,664--88\,688, 2020.

\bibitem{CheCheDas22}
C.~Chen, X.~Chen, D.~Das, D.~Akhmetov, and C.~Cordeiro, ``Overview and
  performance evaluation of {Wi-Fi 7},'' \emph{IEEE Communications Standards
  Magazine}, vol.~6, no.~2, pp. 12--18, 2022.

\bibitem{GalGerCar2023}
{L. Galati-Giordano}, G.~Geraci, M.~Carrascosa, and B.~Bellalta, ``What will
  {Wi-Fi} 8 be? {A} primer on {IEEE} 802.11bn {Ultra High Reliability},''
  \emph{arXiv:2303.10442}, 2023.

\bibitem{ResCor22}
E.~Reshef and C.~Cordeiro, ``Future directions for {Wi-Fi 8} and beyond,''
  \emph{IEEE Communications Magazine}, pp. 1--7, 2022.

\bibitem{OugGerPol2023}
E.~Oughton, G.~Geraci, M.~Polese, and V.~Shah, ``Prospective evaluation of next
  generation wireless broadband technologies: {6G} versus {Wi-Fi} 7/8,''
  \emph{Available at SSRN: https://ssrn.com/abstract=4528119}, 2023.

\bibitem{bellalta2016ieee}
B.~Bellalta, ``{IEEE 802.11ax: High-efficiency WLANs},'' \emph{IEEE Wireless
  Communications}, vol.~23, no.~1, pp. 38--46, 2016.

\bibitem{lopez2019ieee}
D.~L{\'o}pez-P{\'e}rez, A.~Garcia-Rodriguez, L.~Galati-Giordano, M.~Kasslin,
  and K.~Doppler, ``{IEEE 802.11be extremely high throughput: The next
  generation of Wi-Fi technology beyond 802.11ax},'' \emph{IEEE Communications
  Magazine}, vol.~57, no.~9, pp. 113--119, 2019.

\bibitem{yang2020survey}
M.~Yang and B.~Li, ``{Survey and perspective on extremely high throughput (EHT)
  WLAN—IEEE 802.11be},'' \emph{{Mobile Networks and Applications}}, vol.~25,
  no.~5, pp. 1765--1780, 2020.

\bibitem{deng2020ieee}
C.~Deng, X.~Fang, X.~Han, X.~Wang, L.~Yan, R.~He, Y.~Long, and Y.~Guo, ``{IEEE
  802.11be Wi-Fi 7: New challenges and opportunities},'' \emph{IEEE Commun.
  Surveys Tuts.}, vol.~22, no.~4, pp. 2136--2166, 2020.

\bibitem{CarGerKni2023}
M.~Carrascosa, G.~Geraci, E.~Knightly, and B.~Bellalta, ``{Wi-Fi} multi-link
  operation: An experimental study of latency and throughput,'' \emph{IEEE/ACM
  Trans. Networking}, 2023.

\bibitem{mapc}
``{IEEE802.11-22/1512r, UHR Multi-AP coordination and residential Wi-Fi},''
  2022.

\bibitem{wilhelmi2021spatial}
F.~Wilhelmi, S.~Barrachina-Mu{\~n}oz, C.~Cano, I.~Selinis, and B.~Bellalta,
  ``{Spatial reuse in IEEE 802.11ax WLANs},'' \emph{Computer Communications},
  vol. 170, pp. 65--83, 2021.

\bibitem{wilhelmi2019performance}
F.~Wilhelmi, S.~Barrachina-Mu{\~n}oz, and B.~Bellalta, ``On the performance of
  the spatial reuse operation in {IEEE 802.11ax WLANs},'' in \emph{Proc. IEEE
  CSCN}, 2019.

\bibitem{csr_recap}
``{IEEE802.11-22/1822r0}, {Recap} on coordinated spatial reuse operation,''
  2022.

\bibitem{csr_huawei}
``{IEEE802.11-23/0325r0}, {Coordinated} spatial reuse for {UHR},'' 2023.

\bibitem{mentorBroadcom_0855r1}
``{IEEE 802.11-22/0776r1}, {Performance} of {C-BF} and {C-SR},'' 2023.

\bibitem{nunez2022txop}
D.~Nuñez, F.~Wilhelmi, S.~Avallone, M.~Smith, and B.~Bellalta, ``{TXOP sharing
  with coordinated spatial reuse in multi-AP cooperative IEEE 802.11be
  WLANs},'' in \emph{Proc. IEEE CCNC}, 2022.

\bibitem{nguyen2007stochastic}
H.~Q. Nguyen, F.~Baccelli, and D.~Kofman, ``A stochastic geometry analysis of
  dense {IEEE} 802.11 networks,'' in \emph{Proc. IEEE INFOCOM}, 2007.

\bibitem{BelCarGal2023analysis}
B.~Bellalta, M.~Carrascosa, L.~Galati-Giordano, and G.~Geraci, ``Delay analysis
  of {IEEE 802.11be} multi-link operation under finite load,'' \emph{IEEE
  Wireless Communications Letters}, 2023.

\bibitem{bianchi2000performance}
G.~Bianchi, ``{Performance analysis of the IEEE 802.11 distributed coordination
  function},'' \emph{IEEE J. Sel. Areas Commun.}, vol.~18, no.~3, pp. 535--547,
  2000.

\bibitem{boorstyn1987throughput}
R.~Boorstyn, A.~Kershenbaum, B.~Maglaris, and V.~Sahin, ``{Throughput analysis
  in multihop CSMA packet radio networks},'' \emph{IEEE Trans. Commun.},
  vol.~35, no.~3, pp. 267--274, 1987.

\bibitem{ghaboosi2008modeling}
K.~Ghaboosi, B.~H. Khalaj, Y.~Xiao, and M.~Latva-aho, ``Modeling {IEEE} 802.11
  {DCF} using parallel space--time {Markov} chain,'' \emph{IEEE Trans. Veh.
  Tech.}, vol.~57, no.~4, pp. 2404--2413, 2008.

\bibitem{nardelli2012closed}
B.~Nardelli and E.~W. Knightly, ``Closed-form throughput expressions for csma
  networks with collisions and hidden terminals,'' in \emph{Proc. IEEE
  INFOCOM}, 2012.

\bibitem{bellalta2016throughput}
B.~Bellalta, ``{Throughput analysis in high density WLANs},'' \emph{IEEE
  Communications Letters}, vol.~21, no.~3, pp. 592--595, 2016.

\bibitem{barrachina2019dynamic}
S.~Barrachina, F.~Wilhelmi, and B.~Bellalta, ``{Dynamic channel bonding in
  spatially distributed high-density WLANs},'' \emph{IEEE Trans. Mobile
  Computing}, vol.~19, no.~4, pp. 821--835, 2019.

\bibitem{barrachina2019overlap}
------, ``{To overlap or not to overlap: Enabling channel bonding in
  high-density WLANs},'' \emph{Computer Networks}, vol. 152, pp. 40--53, 2019.

\bibitem{barrachina2019komondor}
S.~Barrachina, F.~Wilhelmi, I.~Selinis, and B.~Bellalta, ``{Komondor: A
  wireless network simulator for next-generation high-density WLANs},'' in
  \emph{Proc. IEEE Wireless Days}, 2019.

\bibitem{tgbe_compendium}
``{IEEE 802.11-20/1935r66, Compendium of straw polls and potential changes to
  the Specification Framework Document},'' 2022.

\bibitem{mentorLG_0854r0}
``{IEEE 802.11-23/0854r0}, {Obtaining} {OBSS} {AP} channel information for
  multi-{AP} operation,'' 2023.

\end{thebibliography}

\end{document}